\documentclass[a4paper,twocolumn]{esapub_2}
\usepackage{times}
\usepackage{epsf}
\usepackage[sort&compress,numbers,square]{natbib}
\title{Contemporary interstellar meteoroids in the solar system - in situ
measurements and clues on composition}
\author{M. Landgraf}
\affil{ESA/ESOC, 64293 Darmstadt}
\begin{document}
\maketitle
\keywords{dust; interstellar; in-situ measurements}
\begin{abstract}

Meteoroids originating from the local interstellar medium, traverse
the solar system. This has been proven by in situ measurements by
interplanetary spacecraft as well as highly sensitive radar
measurements. Early attempts to detect interstellar meteoroids using
the instruments on board the Pioneer 8 and 9 spacecraft failed. More
sensitive detectors on board the joint ESA/NASA mission Ulysses as
well as on board the NASA spacecraft Galileo, however, unambiguously
detected meteoroids of interstellar origin. This discovery has started
efforts to compare the results from the in situ measurements with
highly sophisticated models of interstellar dust properties derived
from astronomical absorption and extinction measurements. It was found
that, at least locally, is more mass locked up in meteoroids than
expected from the astronomical measurements. So far the in situ
measurements only allow to derive composition information indirectly
via the meteoroid's dynamics.
\end{abstract}

\section{Sources of Meteoroids in Interplanetary Space}

Our Solar System is filled with small solid fragments, meteoroids,
that originate mainly from larger objects \citep{whipple67}. Asteroids
collide with each other and produce large amounts of meteoroids
\citep[e.g.,][]{grogan00a}. Existing meteoroids impact larger bodies and create
more meteoroids. Comets disintegrate due to solar heating, releasing
gas as well as solid fragments \citep[e.g.,][]{cremonese97}. But not only Solar
System objects are sources of solid particles in the vicinity of the
Sun. Meteoroids, that are known to exist in our galaxy
\citep{mathis90}, enter the Solar System from its local interstellar
environment, and traverse it on unbound trajectories
\citep{zook75b}. Depending on their size, they interact differently
with the gravity and electromagnetic fields of the heliosphere. Table
\ref{tab_sources} gives an overview of the sources of meteoroids in
interplanetary space.
\begin{table}[ht]
\centering
\caption{\label{tab_sources} Sources of meteoroids in interplanetary
space.}
\begin{tabular}{lll}
\hline
source & \multicolumn{2}{c}{characteristics} \\
 & orbit & particle\\
\hline
Asteroids & low eccentricity, & silicate-type, \\
& low inclination & compact \\
Comets & high eccentricity, & carbonaceous, \\
& random inclina- & fluffy\\
& tion & \\
Edgeworth- & low eccentricity, & unknown \\
Kuiper belt & low inclination & \\
Interstellar & hyperbolic & unknown\\
medium & & \\
\hline
\end{tabular}
\end{table}

In this work an overview is given of the in situ measurements of
meteoroids originating from the interstellar medium. Also, first hints
regarding their composition are discussed. The advantage of in situ
measurements using detectors on board interplanetary spacecraft is
that the object is analysed without the contaminating effects of the
Earth. The meteoroids' orbits are preserved until just before impact,
and can, theoretically, be determined with infinite precision. There
are no changes to the shape or chemistry of the meteoroid due to an
entry into an atmosphere. However, because of the limited size and
mass of in situ detectors, so far only rough orbit determination has
been possible and the area of chemical analysis by mass spectroscopy
has just started \citep{srama97,brownlee00}. While the comparison of
the in situ results with isotopically anomalous meteoritic inclusion
is interesting, a full discussion on the latter is out of the scope of
this work. Also, the solid matter that is observed by astronomical
means in cold molecular clouds of our and other galaxies is not
discussed in detail here, despite the fact that it is obviously
connected to the local interstellar meteoroid population.

\section{Dust in the Local Interstellar Medium}
\begin{figure}[t]
\epsfxsize=\hsize
\epsfbox[50 20 500 360]{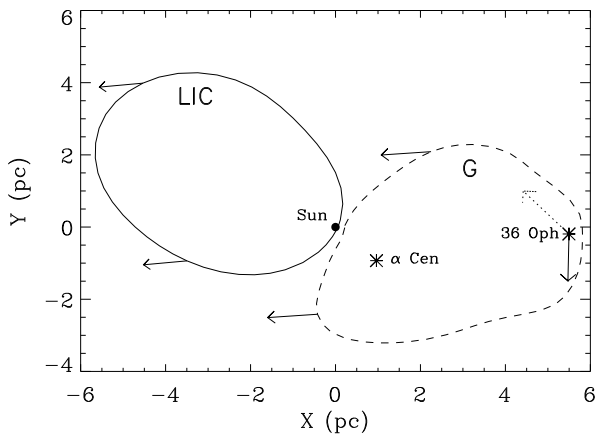}
\caption{\label{fig_lism} Sketch of clouds in the local interstellar
medium as shown by \citet{wood00} in their figure 2. A $10\times
10\:{\rm AU}$ region of the galactic plane is shown with arrows
indicating the heliocentric velocities of the clouds. The local
interstellar cloud (LIC), at the edge of which the Sun is located, is
indicated by the solid line boundary. According to the Wood et
al. model, the Sun will leave the LIC in the near future and will
enter the G-cloud (indicated by the dashed line). The nearby star 36
Oph has a proper motion such that its velocity vector relative to the
G-cloud (dotted line), and thus its astrosphere, has an angle with the
line of sight towards the Sun.}
\end{figure}

The local interstellar medium (LISM) is immersed in a bubble of very
hot ($10^6\:{\rm K}$) and very tenuous ($10^{-3}\:{\rm atoms}\:{\rm
cm}^{-3}$) galactic matter \citep{frisch95,frisch96}. Numerous clouds
are present in this bubble, one of which is the local interstellar
cloud \citep[LIC,][]{lallement92} that surrounds the Sun. Another example
is the G-cloud in which the nearby star $\alpha$ Cen is located. It is
believed \citep{wood00} that the Sun will soon leave the LIC and enter
the G-cloud. Figure \ref{fig_lism} shows the situation.

All material that reaches the Solar System from interstellar space is
per definition part of the LIC. The properties of the LIC are
summarised in table \ref{tab_lic}. The LIC is a typical warm,
partially ionised diffuse cloud \citep{holzer89}. In the diffuse
clouds solid material is normally subject to destruction processes,
returning condensible elements like carbon, oxygen, nitrogen, as well
as silicate, iron, magnesium etc. to the gas phase \citep{jones94}. The
analysis of the elementary composition of the LIC \citep{gry95} shows
that the mass fraction of solid to gaseous material should be between
$1:500$ and $1:400$ \citep{frisch99}.
\begin{table}[ht]
\centering
\caption{\label{tab_lic} Properties of the Local Interstellar Cloud
(LIC) according to \citet{frisch99}.}
\vspace{2mm}
\begin{tabular}{ll}
\hline
\hline
neutral Hydrogen density & $0.22\:{\rm cm}^{-3}$ \\
ionised Hydrogen density & $0.10\:{\rm cm}^{-3}$ \\
motion relative to the Sun & $\lambda = 74.7^\circ$\\
\hspace{2mm} (ecliptic coordinates) & $\beta = -4.6^\circ$ \\
temperature & $6900\:{\rm K}$ \\
magnetic field & few ${\rm \mu}\:{\rm G}$\\
average mass density & $2\times 10^{25}\:{\rm cm}^{-3}$\\
\hline
\end{tabular}
\end{table}
\begin{figure}[t]
\epsfxsize=\hsize
\epsfbox{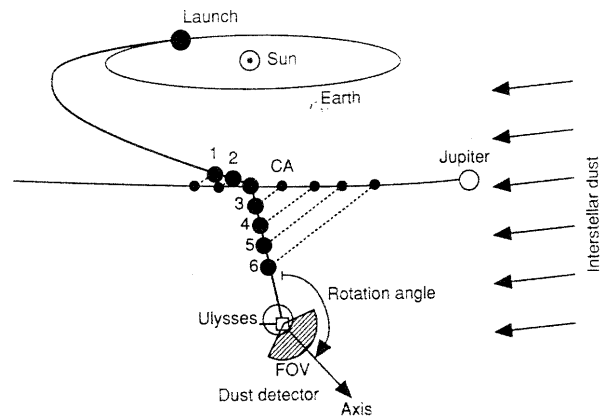}
\caption{\label{fig_ulsflyby} Ulysses measurements of the stream of
interstellar meteoroids (entering from the right). A view from above
the ecliptic plane is shown with the orbits of the Earth, Jupiter, and
the Ulysses spacecraft. The big labelled dots indicate the positions
of Ulysses during measurements of dust streams emanating from the
Jovian system before and after the fly-by. The field of view (FOV) of
the meteoroid sensor on board Ulysses is indicated by the hashed
area. The instrument rotates about the spacecraft's spin axis and thus
the impact rotation angle for each detected meteoroid can be
measured. Most impacts after Jupiter fly-by (disregarding Jovian dust
streams) have been detected at rotation angles around
$90^\circ$. Figure taken from \citet{gruen93}.}
\end{figure}

\begin{figure}[t]
\epsfxsize=\hsize
\epsfbox{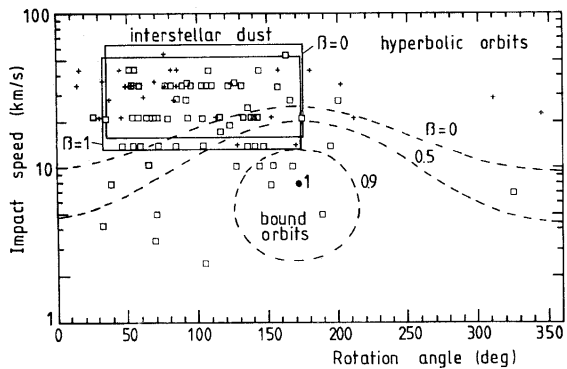}
\caption{\label{fig_velrot} Impact speed and rotation angle of
meteoroids detected by Ulysses after Jupiter fly-by. Crosses mark
impact speed and rotation angle of meteoroids with masses below
$10^{-13}\:{\rm g}$, squares represent meteoroids with masses above
$10^{-13}\:{\rm g}$. The two boxes indicate the uncertainties in the
measurement of the impact direction and velocity around the expected
values for interstellar meteoroids of $26\:{\rm km}\:{\rm s}^{-1}$ and
$100^\circ$ for two values of radiation pressure coefficient
$\beta$. The dashed lines indicate the boundary between bound and
unbound orbits, for different values of $\beta$, where $\beta=0$ means
no radiation pressure, and $\beta=1$ means radiation pressure equals
gravity. Figure taken from \citet{gruen94}.}
\end{figure}

The dust content of more remote interstellar clouds can be determined
by measurements of the extinction of starlight passing through the
cloud \citep{mathis90}. Typically a wavelength-dependent extinction is
observed, caused by solid particles smaller than the wavelengths of
the optical band. The concentration of larger grains can however not
be determined this way, because they absorb and reflect light
independent of the wavelength. The lines of sight through the LIC are
too short to exhibit significant extinction. The emission of infrared
radiation is also below any instrument's sensitivity. It is therefore
impossible to observe solid particles in the LIC directly using remote
sensing.

\section{The Existence of Local Interstellar Meteoroids}
Whether or not local interstellar meteoroids exist and whether they
penetrate the solar system was extensively discussed in the 1970ies
and 1980ies. The Earth orbiting Pioneer 8 and 9 spacecraft that
carried instruments for the detection of impacts by meteoroids did not
find any unambiguously interstellar impactors. It was concluded
\citep{mcdonnell75} that less than $4\%$ of the meteoroids found at
$1\:{\rm AU}$ can be of interstellar origin. This was explained by
modelling efforts \citep{morfill79a,gustafson79} that showed that the
heliospheric magnetic field was able to divert solid particles, as
long as their are not larger than $0.1\:{\rm \mu m}$, which was then
believed to be the upper size limit of interstellar meteoroids
\citep{mathis77} in the diffuse interstellar medium. At that point it
seemed to be clear that no interstellar meteoroids can make it into
the Solar System.

An argument derived from a work by \citet{holzer89} revived the
discussion about the existence of solid interstellar matter in the
Solar System: On average $1\%$ of the mass of the interstellar medium
is contained in dust. If that is true for the LIC, with its average
mass density of $2\times 10^{-25}\:{\rm g}\:{\rm cm}^{-3}$ and its
relative velocity of $26\:{\rm km}\:{\rm }^{-1}$ with respect to the
Sun, an interstellar meteoroid flux density of $10^{-3}\:{\rm
m}^{-2}\:{\rm s}^{-1}$ can be expected. However, the flux at Earth of
meteoroids with masses greater than $10^{-13}\:{\rm g}$ was found to
be $1\times 10^{-4}\:{\rm m}^{-2}\:{\rm s}^{-1}$
\citep{gruen85}, one order of magnitude lower than the value derived
from the theoretical mass density of the LIC. The discussion on the
existence of interstellar meteoroids was settled by the unambiguous
detection of small interstellar particles (masses if typically
$10^{-13}\:{\rm g}$) by an instrument on board the Ulysses spacecraft
after its fly-by of Jupiter \citep{gruen93}.

The evidence of the interstellar origin of the meteoroids detected by
Ulysses comes from three independent observations: (a) The majority of
impacts after Jupiter fly-by came from a retrograde direction,
opposite to the direction of motion of meteoroids from Asteroids and
short-period comets, (b) the impact velocity of particles from the
retrograde direction was higher than the local solar system escape
velocity, even if radiation pressure effects are neglected, and (c)
the rate of impacts from that direction stayed nearly constant over a
range of ecliptic latitude from $0^\circ$ to $79^\circ$. Observation
(a) is illustrated in figure \ref{fig_ulsflyby}, showing the geometry
of the measurements after the fly-by of Jupiter. Most meteoroids were
detected at rotation angles around $90^\circ$. This means that the
observed meteoroids can only be of Solar System origin if a majority
of them moves on retrograde orbits, which contradicts meteor
observations. Observation (b) shows that the orbits of the detected
meteoroids have to be hyperbolic. Due to the uncertainty in the
measurement of the impact velocity, this statement does not hold for
individual impacts, but for the statistic ensemble of impacts measured
from the retrograde direction after Jupiter fly-by (see figure
\ref{fig_velrot}).

\begin{figure}[t]
\epsfxsize=\hsize
\epsfbox{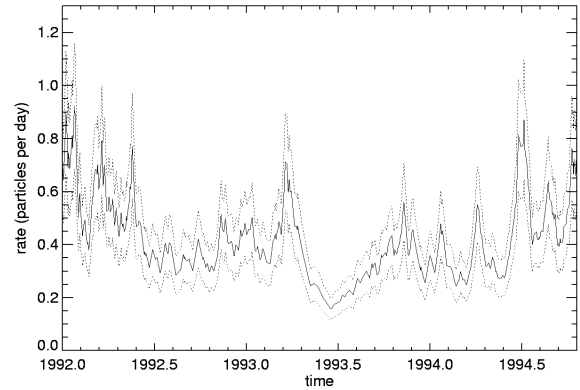}
\caption{\label{fig_rate} Impact rate of meteoroids measured by the
detector on board the Ulysses spacecraft from 1992 to end-1994. A
sliding mean average over $4$ impacts is shown (solid line), together
with $1\sigma$ uncertainties (dotted lines). Ulysses left the ecliptic
plane at $1992.2$ towards increasing ecliptic latitudes until
end-1994. Figure taken from \citet{landgraf96}} 
\end{figure}

A stream of interstellar meteoroids penetrates the whole Solar System
and must thus be detectable independent of ecliptic
latitude. Observation (c) shows that this is true for the Ulysses
observations. The impact rate from all impacts (interstellar plus
interplanetary) is shown in figure \ref{fig_rate}. While it is true
that the rate decreases by less than a factor of $3$ after the
spacecraft leaves the ecliptic plane (February 1992), it levels off
above $0.3$ impacts per day. If only meteoroids of Solar System origin
were detected, the decrease in impact rate would have decreased by
more than an order of magnitude. In combination, observations (a),
(b), and (c) are considered evidence for the existence of interstellar
meteoroids in the Solar System.

\section{Characteristics of the Local Interstellar Dust Population}

The Ulysses spacecraft continues to monitor the stream of interstellar
meteoroids. From these observations three discoveries have been made:
(1) the particle mass distribution is not cut off at $10^{-13}\:{\rm
g}$ as expected from extinction measurements \citep{landgraf00a},
which leads to an apparent excess in dust mass and thus the dust-to
gas ratio of the LIC by at least a factor of $2$, (2) solar radiation
pressure does affect the stream inside $4\:{\rm AU}$
\citep{landgraf99c}, and (3) the flux of small interstellar meteoroids
is modulated with the solar cycle \citep{landgraf00}. In what follows
the three phenomena and their consequences are discussed.

\begin{figure}[t]
\epsfxsize=\hsize
\epsfbox{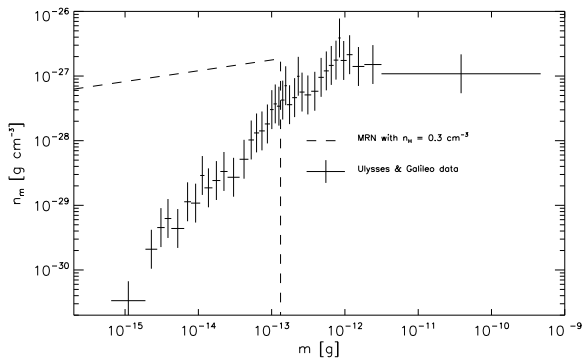}
\caption{\label{fig_massdens} Mass density-mass distribution of
interstellar meteoroids. The crosses represent the combined in situ
measurements by the detectors on board the Ulysses and Galileo
spacecraft. The dashed line shows the particle mass distribution of an
model of the interstellar particle population derived from extinction
measurements as well as cosmic abundance considerations
\citep[MRN,][]{mathis77}. For this model it is assumed that the average
gas density in the LIC is $0.3\:{\rm cm}^{-3}$.}
\end{figure}

The mass density-mass distribution of the interstellar meteoroids
measured by the the Ulysses detector as well as an identical
instrument on board the Galileo spacecraft is shown in figure
\ref{fig_massdens}. The comparison of this local distribution with the
average distribution derived from extinction
measurements \citep{mathis77} over long interstellar lines of sight
contains three interesting observations. First, the mass density
contained small interstellar meteoroids is lower in the local
distribution than in the average interstellar distribution. This was
attributed to the effect of the solar wind magnetic field that
deflects small interstellar meteoroids from the solar system
\citep{landgraf00}. Second, the mass range of interstellar meteoroids
detected locally extends almost $4$ orders of magnitude to higher
masses than was expected from the average extinction measurements
\citep{gruen00}. Since large particles with masses above
$10^{-13}\:{\rm g}$ do not cause a strong wavelength dependent
extinction in the optical and UV bands, the observation of these
grains does not directly contradict the extinction
measurements. However, current models of the interstellar particle
population \citep{mathis96,li97} feature a cut-off to large masses in
the interstellar particle mass distribution, because they are
constrained by the amount of condensible elements available in
interstellar space. This leads to the third observation in figure
\ref{fig_massdens}. The total mass, which is given by the area beneath
the curves in figure
\ref{fig_massdens} locked up in local interstellar meteoroids is by
about a factor of $2$ bigger than expected from cosmic abundance
considerations \citep{frisch99}. Thus, the in situ detection of
relatively large ($m > 10^{-13}\:{\rm g}$) local interstellar
meteoroids apparently violates the cosmic abundance of condensible
elements. It can be speculated that the local interstellar meteoroids are
injected into the LIC by some mechanism and thus do not contribute to
the budget of chemical elements in the LIC. In that case the original
source of local interstellar meteoroid population is still unknown.

\begin{figure}[t]
\epsfxsize=\hsize
\epsfbox{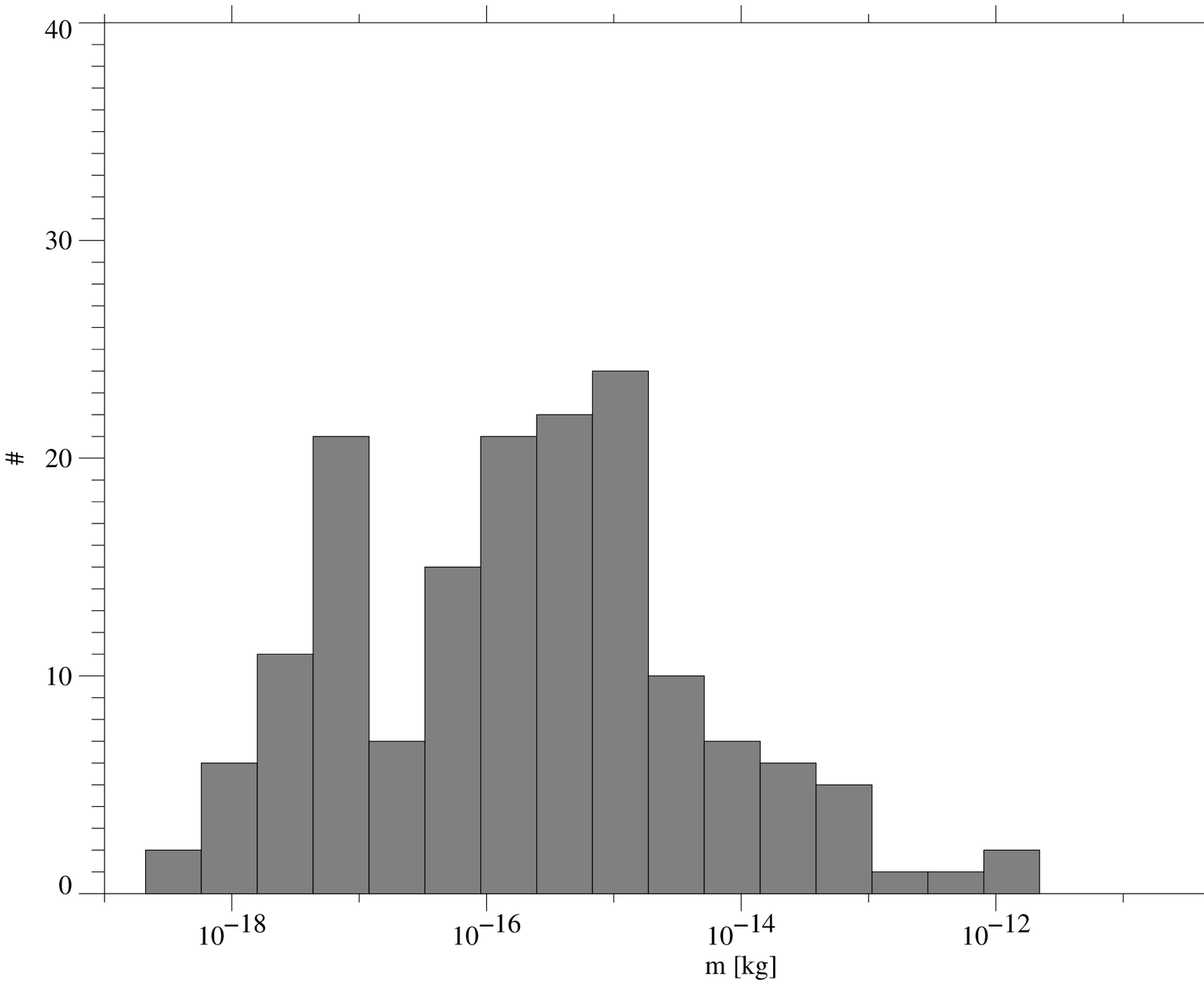}
\epsfxsize=\hsize
\epsfbox{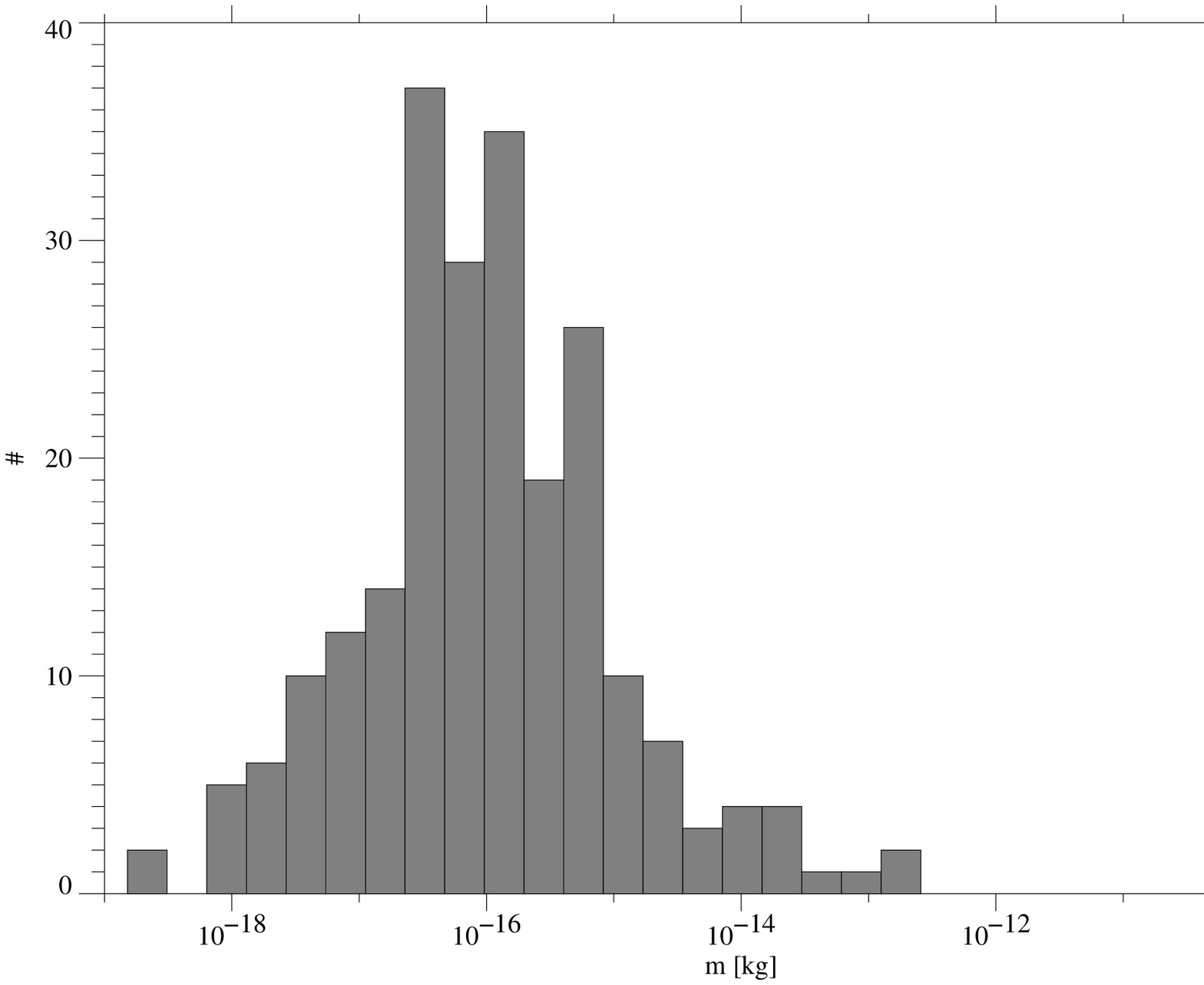}
\caption{\label{fig_betagap} Histograms of the particle mass
distribution measured by the Ulysses detector between $2$ and $4\:{\rm
AU}$ (upper panel), and outside $4\:{\rm AU}$ (lower panel). In the
upper panel a lack of particles between $10^{-17}$ and $10^{-16}\:{\rm
kg}$ can be seen compared to the distribution in the lower panel.}
\end{figure}

The Ulysses measurements of the mass distribution of local
interstellar meteoroids at different heliocentric distances shows that
there is an interaction of the interstellar meteoroid stream with the
solar radiation. Figure \ref{fig_betagap} shows the particle mass
distribution of interstellar meteoroids measured by Ulysses at two
different heliocentric distances. Meteoroids in the mass range between
$10^{-17}$ and $10^{-16}\:{\rm kg}$ are less abundant in the
measurements closer to the sun ($2$ to $4\:{\rm AU}$) than they are in
the measurements more far away from the Sun (outside $4\:{\rm
AU}$). Starting with the assumption that the meteoroid mass
distribution measured outside $4\:{\rm AU}$ is a better representation
of the distribution in the LIC, it can be concluded that some
mass-selective mechanism that is more efficient closer to the Sun
removes the meteoroids in this mass region. The only mechanism that
specifically affects meteoroids in the mass region between $10^{-16}$
and $10^{-17}\:{\rm kg}$ is solar radiation pressure that is most
effective for meteoroids that have sizes in the order of the maximum
wavelength of the solar spectrum ($\lambda_{\rm max}= 450\:{\rm
nm}$). While this observation depends on uncertain micro-physical
properties of the meteoroids such as the bulk mass density and optical
properties, it can be concluded that the material of which the local
interstellar meteoroids consist can neither be too absorbent (like
pure graphite) nor transparent (like pure quartz) \citep[see also the
paper by H. Kimura in this issue]{landgraf99c}.

\begin{figure}[t]
\epsfxsize=\hsize
\epsfbox{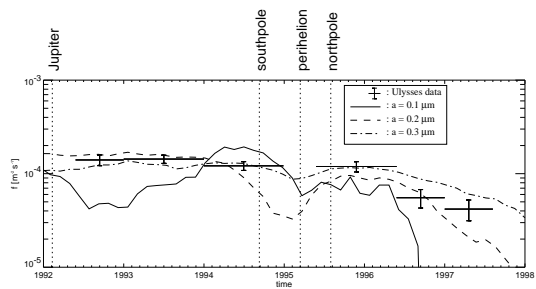}
\caption{\label{fig_ulsflux} Time variability of interstellar
meteoroid flux. The error-bars show the flux measured by the Ulysses
detector and the lines indicate the flux predicted by the magnetic
interaction model for various particle radii. Vertical lines indicate
special events of the Ulysses mission. Figure taken from
\citet{landgraf00}.}
\end{figure}

Another aspect of the Ulysses in situ measurements of interstellar
meteoroids is the time variability of the interstellar stream. After
Jupiter fly-by February 1992 until mid-1996 the interstellar particle
flux appeared to be constant at $1\times 10^{-4}\:{\rm m}^{-2}\:{\rm
s}^{-1}$. Then the flux decreased significantly to $4\times
10^{-5}\:{\rm m}^{-2}\:{\rm s}^{-1}$, about a factor of $3$ below its
former value (see figure \ref{fig_ulsflux}). While it can not be ruled
out that the reason for this decrease is a change in the local
interstellar meteoroid concentration in the LIC, it is more likely due
to the change in the phase of the solar cycle and thus in the
heliocentric magnetic field configuration
\citep{gustafson96a}. Modelling of the interaction of a constant
stream of meteoroids, that acquire an electrostatic charge in
interplanetary space, with the heliocentric magnetic field can account
for this decrease of meteoroid flux in the Solar System. The picture
drawn by the model is the following: the magnetic field polarity
between the solar maxima in 1980 and 1991 was favourable for focusing
the interstellar meteoroid stream towards low heliographic latitudes,
i.e. towards the magnetic equator of the Sun. Starting at the maximum
in 1991, the magnetic field polarity was reversed, causing a
deflecting action of the magnetic field. The average magnetic field
strength at the beginning of a new cycle is not very strong, so the
deflection started to be effective in 1995. Ulysses then consequently
observed a decreasing flux in mid-1996 more than 1 year later due to
the lag caused by the meteoroid's dynamics. The model curves in figure
\ref{fig_ulsflux} fit reasonably well the measured flux values for
spherical particles with radii of $0.2\:{\rm mu m}$, corresponding to
masses of $10^{-16}\:{\rm kg}$, the maximum in the particle mass
distribution.

\section{Future Measurements}
After Ulysses and Galileo a new generation of meteoroid detectors was
launched, one aboard the Cassini spacecraft \citep{srama97} towards
Saturn, and one aboard Stardust, a mission to collect cometary as well
as interstellar meteoroids in situ \citep{brownlee00}. While the
Cassini detector is not expected to find many interstellar meteoroids
due to the orientation of its trajectory downstream of the Sun,
Stardust is specifically designed to collect and measure in situ
interstellar particles. The collected particles will be available for
analysis to the scientific community after the capsule returns to the
Earth in 2006. The Cometary and Interstellar Dust Analyser (CIDA)
instrument on board Stardust has detected few interstellar meteoroids
so far, due to its small sensitive area of $80\:{\rm cm}^2$ and the
reduced interstellar dust flux, as predicted \citep{landgraf99d}. The
information content of these impacts is however high, because CIDA is
capable of determining the elementary composition of the impactors.

The discussion in this paper shows that important information about
our galactic neighbourhood is contained in the stream of interstellar
meteoroids through the Solar System. From what we have learned so far
new questions arise: what is the precise stream direction, and how
close is it to the flow direction of gas from the LIC? Are there other
interstellar meteoroid streams that also pass through the Solar
System, but have lower flux densities so that the Ulysses detector can
not identify them? What is the composition of local interstellar
meteoroids and how does it compare to isotopically anomalous
components of IDPs \citep{bradley94,bradley99,keller00}? In order to
find an answer to these questions new measurements have to be
performed, because the Ulysses dust detector has a limited sensitive
area ($0.1\:{\rm m}^2$) and limited direction, velocity and mass
measurement accuracy. An ideal instrument for these measurements is a
large-area, direction sensing, limited field of view detector as
proposed by \citet[see also the paper by Gr\"un in this issue]{gruen00a}.

\bibliography{dust}
\bibliographystyle{unsrtnat}
\end{document}